\documentstyle[12pt]{article}
\def\fun#1#2{\lower3.6pt\vbox{\baselineskip0pt\lineskip.9pt
  \ialign{$\mathsurround=0pt#1\hfil##\hfil$\crcr#2\crcr\sim\crcr}}}
\newskip\humongous \humongous=0pt plus 1000pt minus 1000pt

\newif\ifdtup

\def\oldreffmt#1{\rlap{[#1]} \hbox to 2\parindent{}}

\def\figfmt#1{\rlap{Figure {#1}} \hbox to 1in{}}

\newcommand {\half} {{\frac {1}{2}}}

\def\beq{\begin{equation}}
\def\eeq{\end{equation}}

\def\bq{\begin{quote}}
\def\eq{\end{quote}}

\relax

\def\lqq{\lq \lq }
\newcommand{\be}{\begin{eqnarray}}
\newcommand{\ee}{\end{eqnarray}}
\begin{document}

\begin{titlepage}
\rightline{Preprint CWRU-MATH}
\rightline{October 1993}
\rightline{hep-th/9310125}
\vskip 1.5truecm
\begin{center}
{\large{\bf Hidden algebra of the $N$-body Calogero problem}}
\vskip 0.5cm
 {\bf Alexander Turbiner}$^{\dagger ,\star}$
\vskip 0.5cm
Math. Dept., Case Western Reserve University, Cleveland, OH 44106
\vskip 0.5cm
\end{center}

\begin{center}
{\large ABSTRACT}
\end{center}
\vskip 0.5 cm
\begin{quote}

A certain generalization of the algebra $gl(N,{\bf R})$ of first-order
differential operators acting on a space of inhomogeneous polynomials
in ${\bf R}^{N-1}$ is constructed. The generators of this (non)Lie algebra
depend on permutation operators. It is shown that the Hamiltonian of the
$N$-body Calogero model can be represented as a second-order polynomial in the
generators of this algebra. Given representation implies that the Calogero
Hamiltonian possesses infinitely-many finite-dimensional invariant subspaces
with explicit bases, which are closely related to the finite-dimensional
representations of above algebra. This representation is
an alternative to the standard representation of the Bargmann-Fock type
in terms of creation and annihilation operators.

\end{quote}

\vfill
\noindent
$^\dagger$On leave of absence from the Institute for Theoretical and
Experimental Physics,
Moscow 117259, Russia\\E-mail: turbiner@cernvm or turbiner@vxcern.cern.ch
\vskip 2mm
\noindent
$^\star$Supported in part by a CAST grant of the US National Academy of
Sciences.

\end{titlepage}

\newpage
The Calogero model \cite{calo} is a quantum mechanical system of $N$
particles on a line interacting via a pairwise potential and defined by
the Hamiltonian
\begin{equation}
\label{e1}
H_{Cal} =\half \sum_{i=1}^N \left[ -d_i^2 + x_i^2 \right] +
\sum_{j < i}^N \frac g {(x_i-x_j)^2} \ \ \ \ \ ,
\end{equation}
where $d_i\equiv \frac \partial {\partial x_i}$, $m$ is the harmonic
oscillator frequency, $g=\nu(\nu -1)$ is the coupling constant.
As it was found by F. Calogero, this model is completely integrable and
possesses some very remarkable properties.
First, in order to be normalizable, all eigenfunctions must  (up to a
certain function as a common factor) be either totally symmetric or totally
antisymmetric.
Secondly, it turned out that the energy spectrum is that of $N$
bosons or fermions interacting via harmonic forces only, but with a
total energy shift proportional to $\nu$ (for a review see \cite{olshper}).
Thus the problem (1) is not only completely integrable, but also
exactly-solvable: the spectrum can be found explicitly, in a closed analytic
form.

Recently,  an extention of the Heisenberg algebra was suggested by
Polychronakos \cite{poly}. A remarkable property of this algebra is that
the `structure constants' depend on the permutation operators
(generators of a certain symmetric group). Using this algebra, the
Bargmann-Fock representation for the Calogero model was constructed
\cite{poly,brink1,brink2}:
the Hamiltonian was rewritten as bilinear combination in generators,
which take the sense of creation and annihilation operators. It allowed
to construct in very simple way the commuting integrals of motion for the
Calogero model \cite{poly} and give an operator solution of this model
deriving an explicit expression for all eigenfunctions for any number
of bodies \cite{brink1,brink2}.

The aim of this Note is twofold. First, we construct an analog of the
irriducible finite-dimensional representations of $gl(N,{\bf R})$ acting on
inhomogeneous polynomials, connected with the above mentioned extended
Heisenberg algebra.
In fact, we arrive at some non-Lie algebras of the first-order
differential operators possessing irreducible finite-dimensional
representations. Their `structure constants' depend on permutation operators.
We denote those algebras by $gl(N,{\bf R})_K$.
Secondly, we show that the transformed $N$-body Calogero Hamiltonian can be
rewritten as a quadratic polynomial in generators of $gl(N,{\bf R})_K$ and
then we look for a hidden algebraic structure of the Calogero model.

Recently, it has been shown \cite{tur}, that all known exactly-solvable
quantum mechanical problems (like the harmonic oscillator, the Coulomb
problem etc.) have two equivalent representations of the Hamiltonian:
(i) well-known Bargmann-Fock representation as a bilinear combination in
creation and annihilation operators
and (ii) a representation as a second-degree polynomial of the generators
of algebra $sl(2,{\bf R})$ of the first-order differential operators on
the line, possessing a finite-dimensional representation in finite-degree
polynomials. We show that the $N$-body Calogero Hamiltonian also has the
second representation:
as a  second-degree polynomial of generators of above-mentioned algebra
$gl(N,{\bf R})_K$.

Let us consider an extended Heisenberg algebra \cite{poly,brink1,brink2}
\footnote{In fact, this algebra already appeared in the paper by E. Wigner
\cite{wig}. This paper was devoted to the problem what is the most general
form of the momentum-coordinate commutator leaving covariant the Heisenberg
equations of motion. Implicitly, E. Wigner also introduced the representation
for the operator $D$ (see below, eq.(9)). I am grateful to Y. Nambu and P.
Freund who paid my attention to the paper \cite{wig}}
\be
\label{e234}
&[D_i \,,x_j ]= A_{ij} =\delta_{ij} (1+\nu \sum_{l=1}^N K_{il})-\nu
K_{ij}, \\ &[ x_i , x_j ] = 0, \\ &[ D_i , D_j ] = 0,
\ee
where $K_{ij}$ are generating elements of the symmetric group,
\be
\label{e56}
&K_{ij}x_j = x_i K_{ij}\ \\ &\quad K_{ij}D_j = D_i K_{ij}\,,\quad
\ee
and
\be
\label{e78}
&K_{ij} = K_{ji}\ ,\ (K_{ji})^2 = I \\ &\quad K_{ij} K_{jl} = K_{jl}K_{li} =
K_{li}K_{ij}
\ee
here all indexes run from 1 to $N$, all quantities $K_{ij}$, $x_l$, and $D_k$
are mutually commuting,
when all the indexes like $i,j,l,$ and $k$ are pairwise noncoinciding. The
parameter $\nu$ in the structure constants $A_{ij}$ is any number, but later
it will be associated with the Calogero coupling constant, see (1). In the
limit
$\nu=0$ the algebra of $D's, x's$ becomes a standard $N$-dimensional Heisenberg
algebra.

 A representation of the `covariant derivative' $D_i$ is unambiguously
fixed by the requirement that $D_i$ induce no new poles when acting on
regular functions of $x_i$ and is given by
\footnote{ The expression for $D_i$ was described by several authors
\cite{dunkl,poly,brink1,brink2} (see also footnote 1). In mathematical
literature, sometimes, this is named "Dunkl operator\lqq. Various forms
of $D_i$ have minor differences and we follow to an expression given
by Brink et al \cite{brink1}}
\be
\label{e9}
 D_i =d_i + \nu \sum_{j\neq i} \frac 1 {(x_i -x_j) } (1-K_{ij} ).
\ee
Here the operators $K_{ij}$ are not used explicitly but rather their
representation on the functions of $x_i$.
This representation (9) of $D_i$  has a crucially important property for our
further consideration: the operators $D_i$ leave the vector space of
polynomials invariant.

Now let us proceed to construct of representations of some algebras using
the extended Heisenberg algebra (2)-(4).
It is rather easy to build up an analog of the vector field representation of
$gl(N)$:
\be
\label{e10}
J_{ij} \ =\ x_i D_j\ , \quad i,j=1,2,\ldots N
\ee
It is easy to see that for generic $\nu$ this algebra possesses an irreducible
finite-dimensional representation given
by homogeneous polynomials in ${\bf R}^N$ like for the non-deformed algebra
 $gl(N,{\bf R})$. We denote this algebra by $gl(N,{\bf R})_K$. The algebra (10)
is not closed.
The commutator of any two generators from (10) is expressed as a linear
combination in (10). However, generically,  those linear combinations contain
the permutation operators $K$'s as coefficients and the commutators of the
r.h.s. and (10) do not belong to the original algebra (10).

We also need a representation of $gl(N,{\bf R})_K$ in terms
of the first-order differential operators acting on functions in
${\bf R}^{N-1}$. In order to derive such a representation, we have to perform
a procedure of so called `projectivization' :
instead of using homogeneous functions in $x \in {\bf R}^N$ as representation
space, one should consider inhomogeneous functions in $y \in {\bf R}^{N-1}$.
Unlike the standard way of introducing the $y$-coordinates:
$y_i= \frac{x_i}{x_N}$ (see e.g. \cite{ams}),
we have to use more sophisticated trick, keeping in mind that we handle to
the $N$-body problem.

First of all, we define Jacobi coordinates as follows
\[ u\ =\ \sum_{p=1}^N x_p\ , \]
\be
\label{e11}
y_i\ =\ x_i - \frac{1}{N} \sum_{j=1}^N x_j\ ,\quad i=1,2,\ldots (N-1)
\ee
where $u$ is a the center-of-mass coordinate. It is evident that, under
such a change of variables, homogeneous functions in $x$ remain as homogeneous
in $(y,u)$. Derivatives in the new coordinates are related with derivatives in
the old coordinates as follows
\[ d_u \equiv \frac{\partial}{\partial u}\ = \ \frac{1}{N} \sum_{p=1}^N d_p
\ ,  \]
\be
\label{e12}
d_{y_i} \equiv \frac{\partial}{\partial y_i}
 \ = \ d_i\ -\ d_N \ ,
\ee
Now let us make a crucial step: perform the procedure of the`projectivization'
(see above) in $(u,y)$--coordinates through introducing new coordinates
\be
\label{e13}
Y_i\ =\ \frac {y_i}{u}\ ,
\ee
and considering the action of generators (10) not on homogeneous functions
in a form $f_x(x_1,x_2,\ldots x_N)$ or
\[
f_y(y_1,y_2,\ldots y_{N-1},u)=u^n f_y(\frac{y_1}{u},\frac{y_2}{u},\ldots
\frac{y_{N-1}}{u}, 1) \equiv u^n \phi (Y_1,Y_2,\ldots Y_{N-1})\ ,
\]
but on inhomogeneous functions $\phi (Y_1,Y_2,\ldots Y_{N-1})$.
Here $n$ is a degree of homogenuity. Without loss of generality, one can set
$u=1$. After a tedious but straightforward calculation, we arrive at the
generators of $gl(N,{\bf R})_K$ algebra in the `projectivized' representation.
They can be presented in the following form $(\partial_i \equiv
\frac{\partial}{\partial Y_i})$ \footnote{For the sake of simplicity,
hereafter we use $`y$' instead of capital letter $`Y$' for the coordinates}
\begin{eqnarray}
\label{e14}
&J_i^- = \partial_i +  \nu \sum_{j\neq i,N} \frac{1}{x_{ij}} (1-K_{ij}) +
\frac{2\nu}{x_{iN}}(1-K_{iN}) +
\nonumber \\
&+\nu \sum_{j\neq i} \frac{1}{x_{jN}} (1-K_{jN})\ ,\quad i=1,2,\ldots (N-1) \ ,
\end{eqnarray}
\be
\label{e15}
J_{ij}^0 = y_i J_j^- \ , \quad i,j=1,2,\ldots (N-1) \ ,
\ee
\be
\label{e16}
J^0 = n - \sum_{p=1}^{N-1} y_p \partial_p \ ,
\ee
\be
\label{e17}
J_i^+ = y_i J^0\ , \quad i=1,2,\ldots (N-1) \ .
\ee
where
\[
x_{ij}=y_i-y_j \ ,\ x_{iN}=y_i+\sum_{p=1}^{N-1} y_p \ .
\]
Let us emphasize that generators (16)-(17) do coincide to those for $\nu=0$
remaining the same as to $gl(N,{\bf R})$.
In the new coordinates (11) the permutation operators act as follows
\begin{eqnarray}
\label{e18}
&K_{ij}y_j \ =\ y_i K_{ij}\ ,\quad j \neq N \ ,
\nonumber
\\
&K_{iN}y_i \ =\ (-\sum_{p=1}^{N-1} y_p) K_{iN}
\end{eqnarray}
and
\begin{eqnarray}
\label{e19}
&K_{ij}\partial_j \ =\  \partial_i K_{ij}
\nonumber \\
&K_{iN} \partial_i \ =\ -\partial_i K_{iN}
\nonumber \\
&K_{iN} \partial_j \ =\ (\partial_j-\partial_i) K_{iN}
\end{eqnarray}
The generators (14)-(17) obey the commutation relations:
\be
\label{e20}
[ J_i^{\pm}, J_j^{\pm}] = 0\ ,
\ee
\be
\label{e21}
[ J_i^{\pm}, J^0] = \pm J_i^{\pm}\ ,
\ee
\be
\label{e22}
[ J_{ij}^0, J^0] = 0\ ,
\ee
\be
\label{e23}
[ J_i^+, J_{j}^-] = J_{ij}^0 - [(\delta_{ij}(1+\nu\sum_{p=1}^N K_{ip})-
\nu(K_{ij}-K_{iN})]J^0\ ,
\ee
\be
\label{e24}
[ J_{ik}^0, J_j^-] = - [(\delta_{ij}(1+\nu\sum_{p=1}^N K_{ip})-
\nu(K_{ij}-K_{iN})]J^-_k \ ,
\ee
\be
\label{e25}
[ J_{ij}^0, J_{k}^+] \ =\ C_{ij,k}^p (K)  J_{p}^+ \ ,
\ee
\be
\label{e26}
[ J_{ij}^0, J_{lm}^0] \ =\ C_{ij,lm}^{pq} (K)  J_{pq}^0 \ .
\ee
where $C_{ij,k}^p (K)$ and $C_{ij,lm}^{pq} (K)$ are rather cumbersome
expressions linear in permutation operators, which become in the limit $\nu$
goes to zero the standard structure constants of the algebra  $gl(N,{\bf R})$.
They will be presented elsewhere.
We say that (20)-(26) defines the algebra  $gl(N,{\bf R})_K$.
In the limit $\nu \rightarrow 0$ the representation (14)-(17) becomes a
degenerate representation of $gl(N,{\bf R})$ (see e.g. \cite{ams}), while
(20)-(26) coincide with the commutation relations of $gl(N,{\bf R})$.

The representation (14)-(17) has an outstanding property: for generic $\nu$,
when $n$ is a non-negative integer number, there appears a finite-dimensional
irreducible representation ${\cal P}_n$ realized on inhomogeneous polynomials
in $y_i,\linebreak i=1,2,\ldots (N-1)$ of degree not higher than $n$. The
dimension of this representation is equal to
\be
\label{e27}
\dim {\cal P}_n = \frac {(n+1)^{N-1}} {2^{N-2}}\ ,\ N\geq 2
\ee
So the representation space is the same for any value of parameter $\nu$.

Now let us proceed to the second part of our program -- a search of the
connections between the algebra of generators (14)-(17) and the Calogero model.
Firstly, let us make a rotation of the Hamiltonian (1) \cite{brink1}
\[
H_{\nu} \equiv {\beta^{-\nu}} H_{Cal} {\beta^{\nu}}=
\]
\[
=\ - \sum_{i=1}^N \left[ d_i^2 + \nu \sum_{j\neq i} \frac 2 {x_i - x_j} d_i
- \nu\sum_{j\neq i} \frac 1 {(x_i-x_j)^2} (1 - K_{ij}) \right] +
\sum_{i=1}^N x_i^2 \ =
\]
\be
\label{e28}
=\ - D^2 + X^2\ ,\quad \beta=\prod_{i>j}(x_i -x_j) \ ,
\ee
where $X^2 = \sum_{i=1}^N x_i^2$, and $g=\nu(\nu\mp 1 )$, where the upper
and lower sign refers to the totally symmetric and antisymmetric
eigenfunctions, respectively. It is worth emphasizing that in (28) there
appear the permutation operators while they do not appear in the original
Hamiltonian. Their occurance reflects the existance of totally symmetric
and totally antisymmetric eigenfunctions only as solutions of the eigenvalue
problem $H_{Cal} \Psi = E\Psi$ \cite{calo}.

In fact, the expression (28) for $H_{\nu}$ can immediately be rewritten
in the form \cite{poly,brink1,brink2}
\be
\label{e29}
H_{\nu}\ =\ \sum_i \{a_i^+ ,a_i^- \} \ ,
\ee
where  $a_i^\mp = (x _i \pm D_i )$ has the meaning of creation/annihilation
operators. This form is nothing but the Bargmann-Fock representation.
It is well known that in this formalism
the general expression for all the eigenfunctions and eigenvalues can be
presented in closed analytic form.

We now go further and make one more rotation of the operator $H_{\nu}$
\[
h\ \equiv\ e^{\frac{X^2}{2}} H_{\nu} e^{-\frac{X^2}{2}}
\]
\[
=\ - \sum_{i=1}^N \left[ d_i^2 - 2x_i d_i + \nu \sum_{j\neq i}
\frac 2 {x_i - x_j} d_i - \nu
     \sum_{j\neq i} \frac 1 {(x_i-x_j)^2} (1 - K_{ij}) \right] +
\]
\be
\label{e30}
+ N + \nu N(N-1)
\ee
Then let us change variables to the center-of-mass coordinates (11)
\[
h \ =\ -N \partial_u^2 + 2u\partial_u - \sum_{i=1}^{N-1} \partial_i^2 +
 \frac{1}{N}(\sum_{i=1}^{N-1} \partial_i)^2
\]
\be
\label{e31}
 +  \sum_{i=1}^{N-1} \left[ 2y_i\partial_i -
\nu \sum_{j\neq i}^N \frac 2 {x_i - x_j} \partial_i + \nu
     \sum_{j\neq i}^N \frac 1 {(x_i-x_j)^2} (1 - K_{ij}) \right]
\ee
Extracting the center-of-mass motion (corresponding to the harmonic
oscillator),
we arrive at the final expression for the operator of relative motion
\[
h_{cm}\ = \ -\sum_{i=1}^{N-1} \partial_i^2 + \frac{1}{N}
(\sum_{i=1}^{N-1} \partial_i)^2 +2 \sum_{i=1}^{N-1} y_i \partial_i
\]
\be
\label{e32}
-2\nu \sum_{i=1}^{N-1} ( \sum_{j\neq i}^N \frac 1 {x_i - x_j}) \partial_i +
\nu \sum_{j\neq i}^N \frac 1 {(x_i-x_j)^2} (1 - K_{ij})
\ee
Then a straightforward calculation leads to the conclusion that (32) can be
written in terms of the generators (14)-(17)
\be
\label{e33}
h_{cm}=-\sum_{i=1}^{N-1}J_i^- J_i^- + \frac{1}{N}\sum_{j,i}^{N-1}J_i^-J_j^-
-2J^0 +2n
\ee
while the value of the parameter $n$ plays no role -- it just changes the
reference point for the energy.

Now we reach the main conclusion of our paper:
\begin{quote}
The Hamiltonian of the Calogero model possesses infinitely-many,
finite-dimensional invariant subspaces with bases in polynomials multiplied
by Gaussian functions. Each subspace is a finite-dimensional
representation of the algebra of the generators (14)-(17). This exposes a
hidden algebraic structure of the Calogero model and explains the
exact-solvability of the model.
\end{quote}

In \cite{ams} it was reported a classification of linear differential
operators having either single or infinite-many, finite-dimensional
invariant subspaces in bases in polynomials, which were named by
{\it quasi-exactly-solvable} and {\it exactly-solvable operators},
respectively. This classification was based on the fact, that some linear
spaces of polynomials are nothing but the representation spaces of irreducible
representation of algebras of the first-order differential operators.
It is rather evident, that this classification can be extended for the case
of the generators (14)-(17). Since the vector space ${\cal P}_n$
of inhomogeneous polynomials in $y_i,\ i=1,2,\ldots (N-1)$ of degree not higher
than $n$ is the irreducible representation space for the algebra of generators
(14)-(17), then any linear differential operator of the order less than $(n+1)$
acting in ${\cal P}_n$ can be presented as a polynomial in generators
(14)-(17).
Basically, this leads to a classification of quasi-exactly-solvable operators
having ${\cal P}_n$ as an invariant subspace.
Since the operators (14)-(17) are graded with respect to the action on
monomials, one can describe the exactly-solvable operator as the operator
having a representation as a polynomial in generators (14)-(17) while this
polynomial has no monomials of positive grading.

Above, we showed that the Calogero model on the line (the rational case
in a notation of \cite{olshper}) is an exactly-solvable
problem in a sense of a definition in \cite{ams} with a hidden algebra
$gl(N,{\bf R})_K$: the r.h.s. in (33) has
no monomials of positive grading. A natural question that
arises concerns what about exactly-solvable Sutherland model on the line
(the trigonometrical case in a notation of \cite{olshper}). Is this model
an exactly-solvable with some hidden algebra like all known exactly-solvable
problems of quantum mechanics ? \footnote{For a discussion of this topic see
\cite{tur}} If the answer is yes, does it have the same hidden
algebra $gl(N,{\bf R})_K$ as the Calogero model?

It is a pleasure to thank D. Gurarie for helpful discussions and the Case
Western Reserve University for kind hospitality extended to me.



\vfill
\begin{thebibliography}{299}
\bibitem{calo}
F. Calogero,
 {\it J. Math. Phys.}, {\bf 10} (1969) 2191, 2197 \\
 {\it J. Math. Phys.}, {\bf 12} (1971) 419

\bibitem{olshper}
M.~A. Olshanetsky and A.~M. Perelomov,
 {\it Phys. Rep.}, {\bf 94} (1983) 313.

\bibitem{poly}
A.~P. Polychronakos,
 {\it Phys. Rev. Lett.}, {\bf 69} (1992) 703.

\bibitem{brink1}
L. Brink, T.~H. Hansson, and M.~A. Vasiliev,
 {\it Phys. Lett.}, {\bf B286} (1992) 109.

\bibitem{brink2}
L. Brink, T.~H. Hansson, S. Konstein and M.~A. Vasiliev,
 {\it Nucl. Phys.}, {\bf B401} (1993) 591.

\bibitem{wig}
E.~P. Wigner,
 {\it Phys. Rev.}, {\bf 77} (1950) 711.

\bibitem{tur}
       A.~V. Turbiner,
{\it Comm. Math. Phys.}, {\bf 118} (1988) 467

\bibitem{dunkl}
O.~A. Chalykh, A.~P. Veselov,
 {\it Comm. Math. Phys.}, {\bf 126} (1990) 611,

C.~F. Dunkl,
 {\it Trans. Amer. Math. Soc.}, {\bf 311} (1989) 167.

\bibitem{ams}
       A.~V. Turbiner, \lqq Lie algebras and linear operators with
       invariant subspace",
       Preprint IHES-92/95 (1992);
       to appear in \lqq Lie Algebras, Cohomologies and New Findings in
       Quantum Mechanics", {\it Contemporary Mathematics}, AMS, 1993,
       N. Kamran and P. Olver (eds.)

\end{thebibliography}
\end{document}